\documentclass[aps,twocolumn,prl,amsmath,amssymb]{revtex4-1}

\usepackage{graphicx}
\usepackage{dcolumn}
\usepackage{bm}
\usepackage{amsmath}
\usepackage{newtxtext}
\usepackage[varg]{newtxmath}
\usepackage[usenames]{color}
\usepackage{natbib}
\usepackage{ulem}

\begin{document}
\title{High-Harmonic Generation in a Correlated Electron System}  
\author{Shohei Imai}
\author{Atsushi Ono}
\author{Sumio Ishihara}
\affiliation{Department of Physics, Tohoku University, Sendai 980-8578, Japan}
\date{\today}
\begin{abstract}  
High-harmonic generation (HHG) in crystalline solids have been examined so far on the basis of one-body energy-band structures arising from 
electron itineracy in a periodic potential. 
Here, we show emergence of HHG signals which are attributed to dynamics of many-body states in a low-dimensional correlated electron system. 
An interacting fermion model and its effective pseudo-spin model on a one-dimensional dimer-type  lattice are analyzed.   
Observed HHG signals in a spontaneously symmetry-broken state, where  charge densities are polarized inside of dimer units, show threshold behavior with respect to light amplitude and 
are interpreted in terms of tunneling and recombination of kink-antikink excitations in an electric field. 
\end{abstract}

\maketitle
\narrowtext

Photoinduced nonequilibrium electron dynamics in solids have attracted much attention of researchers not only in condensed matter physics but also in optical physics. 
Recent great progress in intense laser pulse, ultrafast time-resolved experimental probes, and theoretical methods for nonequilibrium systems has opened up a new research field of exotic light-induced phenomena.
High-order harmonic generation (HHG) is one of the attractive phenomena induced by intense laser light~\cite{huttner,kruchinin,ghimire2}.  
This is a nonlinear and nonperturbative light-matter coupled phenomenon, and is widely recognized to be utilized to generate the attosecond X-ray laser pulse. 
Studies of HHG have been developed in atom- and molecule-gas systems~\cite{agostini, krausz, gallmann, corkum, lewenstein, ishikawa}, in which the HHG spectra consist of 
a characteristic plateau and cut-off energy. 
This behavior is well explained by the three-step model, i.e., a sequential process of ionization, forced oscillatory motion,  and recombination of electrons in  atomic/molecular potentials~\cite{corkum, lewenstein}.  
In crystalline solids where the atom are aligned periodically, electronic processes involved in HHG are considered on the basis of the Bloch energy bands due to electron itineracy~\cite{huttner, kruchinin, ghimire2, ghimire, hohenleutner, liu,yoshikawa, ndabashimiye,higuchi, hawkins, vampa, mcdonald, wu2, hansen, ikemachi,ikemachi2, ikeda,kruchinin2}.
An extended  three-step model based on the band structure was proposed to explain the characteristic plateau in HHG observed in several crystalline solids~\cite{wu2, ikemachi, hansen}.  

Beyond conventional semiconductors and metals, to which the one-body Bloch-band picture is applicable, 
correlated electron systems have great potentialities of HHG. 
Large energy scale and fast dephasing due to strong electron-electron interactions~\cite{ishikawa,liu,ikemachi2,kruchinin2} and existence of multiple ordered phases are supposed to be great advantages for HHG. 
One example is that a large third-harmonic generation in a perturbative regime which is observed in one-dimensional copper oxides is attributed to an electron-electron interaction effect~\cite{kishida, ogasawara}.
In recent years, HHG has started to be examined in correlated electron systems from a viewpoint of quasi-particle motion~\cite{silva, murakami, dejean, takayoshi, zhu}. 

\begin{figure}[t]
\centering
\includegraphics[width=\columnwidth, clip]{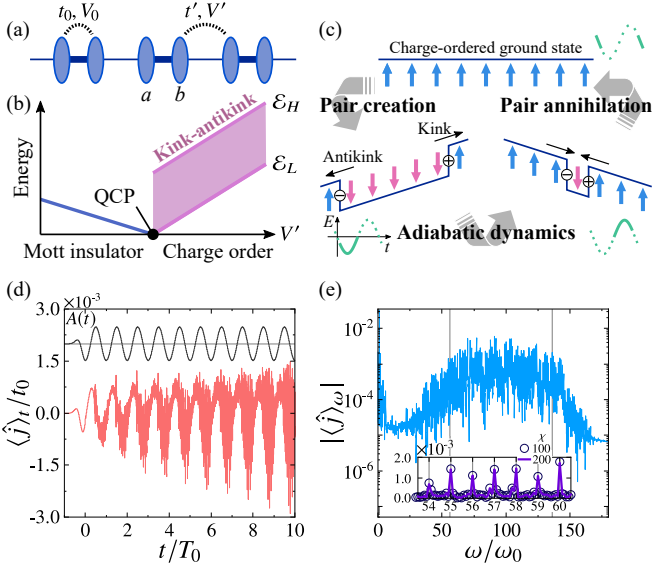}
\caption{
(a) A schematic dimer-type lattice and interactions. The ovals and the thick bars represent atoms and dimer bonds, respectively. 
(b) Schematic kink-antikink excitation spectra in the TI model. 
A shaded area represents the kink-antikink band in the CO state. 
The quantum-critical point (QCP) is indicated as the dot. 
(c) Sketches of the PS configurations, the kink-antikink pair, and the present HHG process. 
(d) Time profiles of vector potential $A(t)$ and current $\langle \hat{j} \rangle_t$. 
(e) Fourier transform of the current $\langle \hat{j} \rangle_\omega$. 
The vertical lines indicate 
${\cal E}_H$ and ${\cal E}_L$. 
Inset shows an enlargement.
The iTEBD method 
is utilized in (d) and (e). 
Inset of (e) shows the results with $\chi=100$ and 200.   
We set $V'/(4t_0)=2.4$, $A_0=5.8$, $\omega_0/t_0=0.1$, and $\tau=1/\omega_0$. 
}
\label{fig:fig1}
\end{figure}

In this Letter, we show that HHG spectra emerge owing to many-electron dynamics in a correlated electron system, rather than the Bloch electron itineracy. 
We analyze photoinduced dynamics of an interacting fermion model on a dimer-type lattice and its low-energy effective model described by the pseudo-spin (PS) operators. 
We find emergence of HHG spectra in a spontaneously symmetry-broken state, in which charge densities are polarized inside of dimer units. 
The HHG spectra show a threshold behavior with respect to light amplitude. 
As shown in Fig.~\ref{fig:fig1}(c), the observed HHG are explained by an extended three-step like processes 
for many-body kink-antikink excitations, which 
are valid even without electron itineracy.

A target system of the present study is an interacting electron system with a one-dimensional dimer-type lattice structure shown in Fig.~\ref{fig:fig1}(a). 
It is well known that when the average number of electrons is 0.5 per site under a strong on-site electron interaction,  there are two competing electronic states in the ground state (GS): a Mott insulating state where the bonding-orbital band is half filled, termed the dimer Mott (DM) insulating state, and the polar charge-ordered (CO) state where electron distribution inside the each dimer unit breaks the inversion symmetry [see Fig.~\ref{fig:fig1}(c)]~\cite{kino,naka,hotta}. 
We analyze the two model Hamiltonians introduced below~\cite{MM}. 
An interacting spinless-fermion model in a dimer lattice is defined as  
\begin{align}
{\cal H}_{\rm SF}&=
-\sum_{i}  \left (t_0  f_{i a }^\dagger f_{i b }^{}+ {\rm H.c.} \right )
-\sum_{i } \left (t' f_{i b }^\dagger f_{i+1 a }^{}+ {\rm H.c.} \right ) 
\nonumber \\
&\quad +V_0\sum_i n_{i a} n_{i b}+V'\sum_{i}  n_{i b} n_{i+1 a} , 
\label{eq:sfH}
\end{align}
where $f_{i \gamma}^\dagger$ ($f_{i \gamma}$) is the creation (annihilation) operator of a spinless fermion at the $i$th unit cell and sublattice $\gamma\ (=a, b)$, and $n_{i \gamma}=f_{i \gamma}^\dagger f_{i\gamma}$ is the number operator. 
The first two terms represent the fermion hoppings, and the last two terms describe the inter-site Coulomb interactions. 
The total number of the fermions is set to $N$ with $N$ being the total number of the dimer units.  
%
We analyze another Hamiltonian for an interacting PS system as a low-energy effective model of ${\cal H}_{\rm SF}$~\cite{naka, tsuchiizu}. 
The local electronic states inside the dimer unit are described by the PS operator: the up and down PSs  imply the states where electron occupies the $a$ (left) and $b$ (right) sites in the dimer unit, respectively. 
The low-energy physics is mapped onto the transverse Ising (TI) model defined by 
\begin{align}
{\cal H}_{\rm TI}=-\frac{V'}{4}\sum_{\langle ij \rangle} { \sigma}_i^z {\sigma}_j^z
-t_0 \sum_i \sigma_i^x , 
\label{eq:TI}
\end{align}
where ${\bm \sigma}_i$ are the Pauli matrices located at the $i$th unit cell. 
The first term (${\cal H}_I$) and the second term (${\cal H}_T$) describe the interaction between the nearest neighbor unit cells, and the transverse field, respectively. 
These terms correspond to the inter-dimer Coulomb interaction and the intra-dimer hopping in Eq.~(\ref{eq:sfH}), respectively.  
%
This model is suitable to study the collective excitations, i.e., the kink and antikink (domain-wall) excitations. 

A vector potential of light is introduced as the Peierls phase as  
$t_0 \rightarrow t_0 e^{-iA(t)}$ and $t' \rightarrow t' e^{-iA(t)}$ in Eq.~(\ref{eq:sfH}), where $A(t)$ is the vector potential at time $t$ and the difference in the bond lengths are neglected. 
The electric field is given by $E(t)=-\partial A(t)/\partial t$. 
This coupling corresponds to the rotation of the transverse field as follows~\cite{MM}:  
${\cal H}_T$ is replaced by 
$
-t_0 \sum_i 
\left [
\cos A(t)\, \sigma_i^x  -\sin A(t)\, \sigma_i^y
\right ]$, 
and the electric current operator is identified as 
$\hat{j}(t)
=-(t_0/N) \sum_i  
\left [
\sin A(t)\, \sigma_i^x  +\cos A(t)\, \sigma_i^y 
\right ]$. 
We confirmed that the numerical results of HHG in the above two models qualitatively coincide in the polar CO state, and 
we will mainly present the results for the TI model. 

The GS and excited states in the TI model without the light field 
have been settled~\cite{kirilyuk}.
The GS is the DM insulating state (a paramagnetic PS state), i.e., $\langle \sigma^z \rangle= 0$ for $V'/(4t_0) < 1$, and 
is the polar CO state (a ferromagnetic PS state) with spontaneous symmetry breaking of the space-inversion symmetry, i.e., $\langle \sigma^z \rangle \ne 0$ for $V'/(4t_0)>1$. 
The boundary at $V'/(4t_0)=1$ is the quantum critical point. 
In order to calculate the transient dynamics induced by the light field in the thermodynamic limit,  the infinite time-evolving block decimation (iTEBD) method is adopted~\cite{vidal}. 
The second-order Suzuki-Trotter decomposition is utilized to calculate the time-evolution of the wavefunction $|\psi(t+\delta t) \rangle \approx e^{-i{{\cal H}(t)} \delta t}|\psi(t)\rangle$ with a small time difference $\delta t$ and the time-dependent Hamiltonian ${\cal H}(t)$. 
In most of the numerical calculations, the maximum number of the matrix dimension ($\chi$) in the iTEBD method, and the time difference are chosen to $\chi=200$ and $\delta t=0.01/t_0$, respectively, which are enough to obtain well convergent results as shown later [see inset of Fig.~\ref{fig:fig1}(e)]. 
We also adopt the exact diagonalization (ED) method based on the Lanczos algorithm for finite size clusters, where the total number of dimer units is $N=16$ and $18$ 
with the periodic-boundary condition.  
The optical absorption spectra [see Fig.~S.4 in the Supplemental Material (SM)] is schematically depicted in Fig.~\ref{fig:fig1}(b). 
In the polar CO state, the excitation spectra are attributed to the kink-antikink pair excitations, and 
exhibit a continuous band where 
the upper and lower edges of the band are ${\cal E}_H=4(V'/4+t_0)$ and ${\cal E}_L=4(V'/4-t_0)$, respectively. 
In the DM state, the low-energy collective excitation is located at $2(t_0-V'/4)$.

First, we show the HHG spectra in the polar CO state [$V'/(4t_0)=2.4$] in the continuous-wave (cw) light. We set 
$A(t)=-A_0 e^{-t^2/(2\tau^2)} \cos (\omega_0 t) $ for $t<0$, and $A(t)=-A_0 \cos(\omega_0 t)$ for $t>0$ 
with frequency $\omega_0$, amplitude $A_0$, and raising time $\tau$. 
Numerical values of $\omega_0$ are chosen to be much smaller than 
the excitation energy gap $\Delta_{\rm gap}={\cal E}_L$.   
Time profiles of the electric current $\langle \hat{j} \rangle_t$ and its Fourier transform $\langle \hat{j} \rangle_\omega$, as well as $A(t)$, are shown in Figs.~\ref{fig:fig1}(d) and \ref{fig:fig1}(e).  
A multiple pulse-like profile with period of $T_0\equiv 2\pi/\omega_0$ appears in $\langle \hat{j} \rangle_t$, and a series of sharp spikes at $\omega=n\omega_0$ with integer number $n$ appear in $\langle \hat{j} \rangle_\omega$ [see inset of Fig.~\ref{fig:fig1}(e)].  
The HHG spectra show a plateau approximately between 
${\cal E}_L$ and ${\cal E}_H$, indicating the nonperturbative processes for this HHG.  
Owing to the breaking of the space-inversion symmetry in the GS, 
both the odd and even orders of high harmonics emerge. 
Overall features mentioned above do not depend on $\chi(\ge 100)$  and are almost reproduced by the ED method 
in finite-size clusters as shown in Fig.~S.2 in SM. 

\begin{figure}[t]
\centering
\includegraphics[width=\columnwidth, clip]{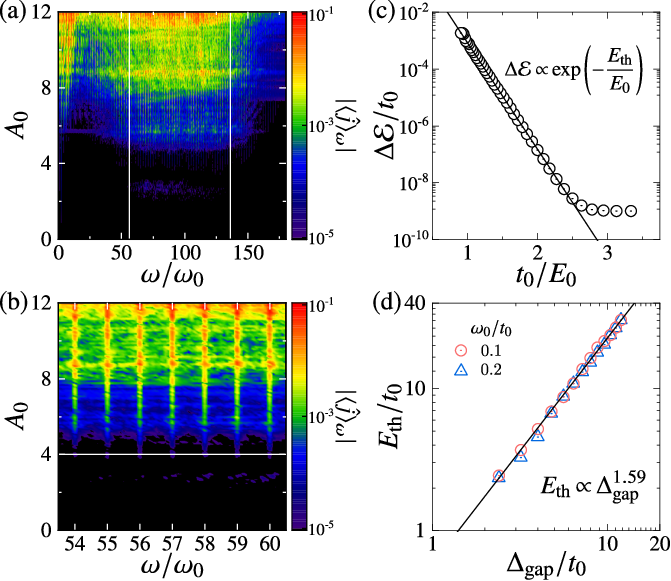}
\caption{
(a) Intensity maps of $\langle \hat{j} \rangle_\omega$.
The vertical lines show the lower and upper edges of the kink-antikink band. 
(b) An enlargement of (a).
The horizontal line indicates threshold intensity.  
(c) A logarithmic plot of the absorbed energy $\Delta {\cal E}$ as a function of inverse of $E_0$. 
The bold line shows $\Delta {\cal E} \propto  \exp(-E_{\rm th}/E_0)$. 
(d) A threshold electric field $E_{\rm th}$ as a function of the gap energy $\Delta_{\rm gap}$. 
The bold line shows $E_{\rm th} \propto \Delta_{\rm gap}^\alpha$ with $\alpha \sim 1.59$. 
The cw and one-cycle pulse field are introduced in (a) and (b), and (c) and (d), respectively. 
We set $V'/(4t_0)=2.4$, $\omega_0/t_0=0.1$, and $\tau=1/\omega_0$. 
The iTEBD method is utilized. 
}
\label{fig:fig2}
\end{figure}
The HHG spectra are sensitive to the light amplitude $A_0$. 
In Figs.~\ref{fig:fig2}(a) and \ref{fig:fig2}(b), the intensity map of $\langle \hat{j} \rangle_\omega$ in the $\omega$-$A_0$ plane and its enlargement are shown, respectively.  
The threshold behavior of the HHG spectra with respect to $A_0$ is clearly seen. 
We find that the threshold decreases with decreasing $V'$ (see Fig.~S.1 in SM). 
\begin{figure}[t]
\centering
\includegraphics[width=\columnwidth, clip]{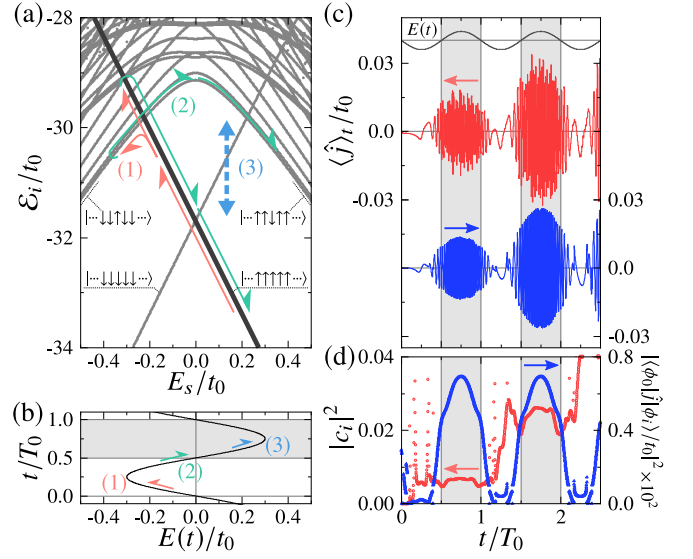}
\caption{
Results based on the adiabatic kink-antikink dynamics.  
(a) Energy level diagram as a function of $E_{s}$. 
The bold line represents the energy of the state $|\phi_0(E_s) \rangle $ adiabatically connected to the all-up state at $E_s=0$.  
(b) A schematic time profile of $E(t)$. 
(c) Time profiles of $E(t)$ (black line) and $\langle \hat{j} \rangle_t$ (red line) induced by the cw light. 
The blue line shows the results calculated from the energy level diagram in (a) (see text).    
(d) Time profiles of the population of the most dominant excited state $|c_i|^2$ (red circles), 
and a square of the transition amplitude $|\langle \phi_0(E_s) |\hat{j}| \phi_i(E_s\rangle |^2$ (blue triangles).
Shaded areas in (b)-(d) represent time domains where $E(t)$ is positive. 
We set $V'/(4t_0)=1.6$, $A_0=3.6$, $\omega_0/t_0=0.1$, $\tau=1/\omega_0$, and $N=18$. 
}
\label{fig:fig3}
\end{figure}

The observed HHG is understood by repetition of dynamics induced by a one-cycle pulse. 
Thus, to reveal the threshold behavior in more detail, 
we examine responses to a one-cycle pulse 
given by $A(t)=-A_0 e^{-t^2/(2\tau^2)} \cos(\omega_0 t)$. 
Using the iTEBD method, we analyze the absorbed energy defined by  
$\Delta {\cal E} \equiv {\cal E}(t \gg \tau)-{\cal E}(t\ll -\tau)$ with total energy ${\cal E}=\langle {\cal H}_{\rm TI} \rangle/N$, which reflects 
population of the excited states induced by the pulse field. 
In Fig.~\ref{fig:fig2}(c), we plot $\Delta {\cal E}$ as a function of $1/E_0$  with the electric field amplitude $E_0\equiv A_0 \omega_0$. 
We note that, instead of the vector potential, a response to the electric field is suitable to examine the breakdown phenomena which will be introduced later. 
The current operator in this case is defined in Eq.~(8) in SM. 
The exponential dependence is observed as 
$\Delta  {\cal E} \propto \exp(-E_{\rm th}/E_0)$ with a threshold electric field $E_{\rm th}$. 
A deviation of data from this function for $\Delta {\cal E}/t_0 < 10^{-9}$ is attributed to the numerical artifact.   
This behavior implies a nonperturbative processes in HHG, and is reproduced by the ED method in finite clusters [see Fig.~S.3(c) in SM]. 
The threshold amplitude calculated in several values of $V'$ is scaled by the excitation energy gap $\Delta_{\rm gap}=4(V'/4-t_0)$ as 
$E_{\rm th} \propto \Delta_{\rm gap}^{\alpha}$ with $\alpha \sim 1.59$ as shown in Fig.~\ref{fig:fig2}(d). 
This indicates a Landau-Zener-like breakdown in the HHG,  except for the exponent which is different from $\alpha=2$ in the case of a static field. 

The observed HHG spectra and their characteristic time profiles are interpreted through the following analysis based on adiabatic kink-antikink dynamics.  
We consider the TI model in an electric field, ${\cal H}_{\rm TI}-[E(t)/2] \sum_i \sigma_{i}^z$, and examine this by using the ED method. 
The energy levels as functions of a static electric field $E(t)=E_s$ is shown in Fig.~\ref{fig:fig3}(a). 
The eigen wavefunction and eigen energy for finite $E_s$ are denoted by 
 $|\phi_i(E_s) \rangle$ and ${\cal E}_i(E_s)$ $(i \geq 0)$,  respectively, 
which are adiabatically connected to the $i$th eigen state at $E_s=0$. 
The GS at $E_s=0$ are doubly degenerated in the thermodynamic limit,  
i.e., the all-up and all-down states, schematically $| \cdots \uparrow \uparrow \uparrow \uparrow \uparrow \cdots \rangle$ and $| \cdots  \downarrow \downarrow \downarrow \downarrow \downarrow \cdots \rangle$, respectively, and the excited states are continuum with the finite excitation gap from GS.
With increasing $E_s$ from zero, the energy of the all-up (all-down) state decreases (increases).  

Then, we examine the current induced by the cw field shown in Fig.~\ref{fig:fig1}(c) and Fig.~\ref{fig:fig3}(c) (red line) from a viewpoint 
of the adiabatic dynamics of many-body states. 
The wavefunction at time $t$ is expanded as 
$|\psi (t) \rangle =\sum_{i \ge 0} c_i |\phi_i (E_s) \rangle \exp[-i {\cal E}_i(E_s)t] $ with coefficients  
$c_i=\langle \phi_i(E_s) |\psi(t) \rangle$. 
Here, we assume that $E_s$ is equal to $E(t)$ and 
$|\phi_0(E_s) \rangle$ is adiabatically connected to the all-up state at $E_s=0$.  
Since $|c_0|^2 \approx 1$ and $|c_{i \ge 1}|^2 \ll 1$  
as shown in Fig.~\ref{fig:fig3}(d), 
the current at time $t$ is approximately given by     
$\langle \hat{j} \rangle_t=\langle \psi(t)| \hat{j}|\psi(t)\rangle$ 
$\approx \sum_{i > 0} c_i c_0^\ast \langle \phi_0(E_s)|\hat{j}|\phi_i(E_s) \rangle \exp[-i\{{\cal E}_i(E_s)-{\cal E}_0( E_s)\}t] +{\rm c.c.} $
In Fig.~\ref{fig:fig3}(c), we compare a time profile of the current calculated by the above method shown by blue line  with that by the real-time evolution. 
We adopt the most dominant excited state among $|\phi_i(E_s) \rangle$'s. 
The two results almost coincide.  
We conclude that this picture based on the adiabatic dynamics is valid to understand the real time processes in the present HHG. 
The facts $|c_0|^2 \approx 1$ and $|c_{i \ge 1 }|^2 \ll 1$ reflect the off-resonant excitation with the light frequency $\omega_0 \ll \Delta_{\rm gap}$, and nonperterpative tunneling processes are incorporated in $c_i$ for $i \ge 1$. 
In the time profile of $\langle \hat{j} \rangle_t$ in Fig.~\ref{fig:fig3}(c), a fine oscillation is attributed to the exponential factor $\exp[-i\{{\cal E}_i(E_s)-{\cal E}_0( E_s)\}t]$. 
An envelope with period of $T_0\ (=2\pi/\omega_0)$, showing large amplitude in regions with positive $E(t)$ [shaded areas in Fig.~\ref{fig:fig3}(c)], is due to the amplitude factor $c_i \langle \phi_0(E_s)|\hat{j}|\phi_i(E_s) \rangle$. 
As shown in Fig.~\ref{fig:fig3}(d), 
this characteristic time profile of the amplitude factor does not originate mainly from populations of the excited state $|c_i|^2$, 
but the transition amplitude $|\langle \phi_0(E_s) |\hat{j}| \phi_i(E_s\rangle |$. 
This means that, in contrast to the HHG in semiconductors, the many body character of the wavefunction governs the transition amplitude in the present case as discussed below.

\begin{figure}[t]
\centering
\includegraphics[width=\columnwidth, clip]{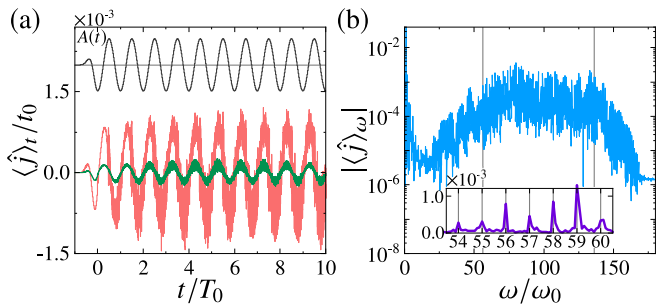}
\caption{
Results in the spinless-fermion model in the cw field.  
(a) Time profiles of $A(t)$ (upper) and $\langle \hat{j} \rangle_t$ [red (light gray) line in lower]. 
The green (dark gray) line in lower represents  $\langle \hat{j} \rangle_t$ where the inter-dimer current is only taken into account. 
(b) Fourier transform of the current, $\langle \hat{j} \rangle_\omega$. 
Inset shows an enlargement. 
We set $V'/(4t_0)=2.4$, $V_0/V'=1$, $t'/t_0=0.5$, $A_0=5.8$, $\omega_0/t_0=0.1$, and $\tau=1/\omega_0$. 
The iTEBD method is utilized. 
}
\label{fig:fig4}
\end{figure}
Now, a microscopic process of the HHG is explained by the adiabatic many-body 
energy diagram in Fig.~\ref{fig:fig3}(a). 
Let us start from the all-up state, $| \cdots \uparrow \uparrow \uparrow \uparrow \uparrow \cdots \rangle$, at $E_s=0$ as a symmetry-broken GS.  
The first excited state at $E_s=0$ is given by a liner combination of the single kink-antikink excitation states, 
and is adiabatically connected to $|\cdots \downarrow \downarrow \uparrow \downarrow \downarrow \cdots \rangle$ in the limit of $E_s \rightarrow - \infty$.
With decreasing $E_s$ from zero, 
the energy level of the all-up state increases and anticrosses with the excited state [(1) in Figs.~\ref{fig:fig3}(a) and (b)]. 
This anticrossing point is known as the quantum spinodal point~\cite{raedt, miyashita}. 
Through the Landau-Zener transition, the state which is adiabatically connected to the all-up state transfers with finite probability to the excited state $|\cdots \downarrow \downarrow \uparrow \downarrow \downarrow \cdots \rangle$ at a certain negative $E_s$. 
However, the transition amplitude $\langle \phi_0(E_s)|\hat{j}|\phi_i(E_s) \rangle$ is almost zero between the all-up state and this excited state, since the current operator  proportional to $\sum_i \sigma_i^y$ brings about one PS flip. 
When a sign of $E(t)$ is turned into positive, the excited state moves adiabatically [(2) in Figs.~\ref{fig:fig3}(a) and (b)], and 
is changed into $|\cdots \uparrow \uparrow \downarrow \uparrow \uparrow \cdots \rangle$ at large positive $E_s$. 
The transition amplitude $\langle \phi_0(E_s)|\hat{j}|\phi_i(E_s) \rangle$ is finite between this excited state and the initial GS, and the system returns to the initial state accompanied by light emission [(3) in Figs.~\ref{fig:fig3}(a) and (b)]. 
This description based on the many-body states corresponds to the three-step model known for the atoms and semiconductors, 
and is valid at least in the region where 
$A_0$ is close to the threshold and $V'/(4t_0) \gtrsim 2.4$.

So far, effects of the electron itineracy is neglected on the HHG spectra calculated in the TI model, where 
the inter-dimer hoppings are not taken into account. 
It is widely recognized that, in Mott insulators as well as  conventional semiconductors the electron itineracy due to the inter-site hopping integral is essential for HHG~\cite{silva,murakami}. 
Beyond the TI model, 
we analyze the HHG in the interacting fermion model given in Eq.~(\ref{eq:sfH}), where the polar CO state is realized for $V' \gg t_0 , t'$.  
A time profile of $\langle \hat{j} \rangle_t$ 
is shown in Fig.~\ref{fig:fig4}(a) where we set $t'=0.5t_0$, and $\langle \hat{j} \rangle_\omega$ is shown in Fig.~\ref{fig:fig4}(b). 
Amplitude of the electric current is dominated by the intra-dimer component, and contribution from the inter-dimer current is less than $10\%$ (see green line in Fig~\ref{fig:fig4}(a)), although amplitude of $t'$ is chosen to be half of $t_0$. 
We conclude that the essential characters in the HHG in the polar CO state is not governed by the electron propagation over the dimer units, but by the kind-antikink excitations and propagations.

In summary, we investigated HHG in the spontaneously symmetry-broken state realized in the TI model 
and the spinless-fermion model as the effective models of the interacting electrons in a dimer-type lattice structure. 
The kink-antikink dynamics are responsible for the present HHG, instead of the electron itineracy. 
The many-body character in the wavefunction governs the transition amplitude between GS and excited states. 
Experimental observations are crucial to confirm the present theoretical prediction for a new mechanism of HHG.  
Candidate materials are low-dimensional organic molecular solids, (TMTTF)$_2$X (TMTTF=Tetramethyltetrathiafulvalene, X=PF$_6$, AsF$_6$), which show the polar CO phase in low temperatures~\cite{javadi,monceau,iwai}. 

\begin{acknowledgements}
The authors would like to thank Y. Masaki, S. Iwai, and M. Naka for their fruitful discussions. 
This work was supported by JSPS KAKENHI, Grant Numbers JP17H02916 and JP18H05208. 
Some of the numerical calculations were performed using the facilities of the Supercomputer Center, the Institute for Solid State Physics, The University of Tokyo.
\end{acknowledgements}




\begin{thebibliography}{99} 








%



\bibitem{huttner}
U.~Huttner, M.~Kira, and S.~W.~Koch,
Laser Photonics Rev.\ \textbf{11}, 1700049 (2017).

\bibitem{kruchinin}
S.~Y.~Kruchinin, F.~Krausz, and V.~S.~Yakovlev,
Rev.\ Mod.\ Phys.\ \textbf{90}, 021002 (2018).

\bibitem{ghimire2}
S.~Ghimire and D.~A.~Reis,
Nat.\ Phys.\ \textbf{15}, 10 (2019).


\bibitem{agostini}
P.~Agostini and L.~F.~DiMauro,
Rep.\ Prog.\ Phys.\ \textbf{67}, 813 (2004).

\bibitem{krausz}
F.~Krausz and M.~Ivanov,
Rev.\ Mod.\ Phys.\ \textbf{81}, 163 (2009).

\bibitem{gallmann}
L.~Gallmann, C.~Cirelli, and U.~Keller,
Annu.\ Rev.\ Phys.\ Chem.\ \textbf{63}, 447 (2012).

\bibitem{corkum}
P.~B.~Corkum,
Phys.\ Rev.\ Lett.\ \textbf{71}, 1994 (1993).

\bibitem{lewenstein}
M.~Lewenstein, P.~Balcou, M.~Y.~Ivanov, A.~L'Huillier, and P.~B.~Corkum,
Phys.\ Rev.\ A \textbf{49}, 2117 (1994).

\bibitem{ishikawa}
K.~L.~Ishikawa and T.~Sato,
IEEE J.\ Sel.\ Top.\ Quantum Electron.\ \textbf{21}, 8700916 (2015).



\bibitem{ghimire}
S.~Ghimire, A.~D.~DiChiara, E.~Sistrunk, P.~Agostini, L.~F.~DiMauro, and D.~A.~Ries,
Nat.\ Phys.\ \textbf{7}, 138 (2011).


\bibitem{hohenleutner}
M.~Hohenleutner, F.~Langer, O.~Schubert, M.~Knorr, U.~Huttner, S.~W.~Koch, M.~Kira, and R.~Huber,
Nature \textbf{523}, 572 (2015).

\bibitem{ndabashimiye}
G.~Ndabashimiye, S.~Ghimire, M.~Wu, D.~A.~Browne, K.~J.~Schafer, M.~B.~Gaarde, and D.~A.~Reis,
Nature \textbf{534}, 520 (2016).


\bibitem{liu}
H.~Liu, Y.~Li, Y.~S.~You, S.~Ghimire, T.~F.~Heinz, and D.~A.~Reis,
Nat.\ Phys.\ \textbf{13}, 262 (2017).

\bibitem{yoshikawa}
N.~Yoshikawa, T.~Tamaya, and K.~Tanaka,
Science \textbf{356}, 736 (2017).

\bibitem{higuchi}
T.~Higuchi, M.~I.~Stockman, and P.~Hommelhoff,
Phys.\ Rev.\ Lett.\ \textbf{113}, 213901 (2014).

\bibitem{hawkins}
P.~G.~Hawkins, M.~Y.~Ivanov, and V.~S.~Yakovlev,
Phys.\ Rev.\ A \textbf{91}, 013405 (2015).

\bibitem{vampa}
G.~Vampa, C.~R.~McDonald, G.~Orlando, P.~B.~Corkum, and T.~Brabec,
Phys.\ Rev.\ B \textbf{91}, 064302 (2015).

\bibitem{mcdonald}
C.~R.~McDonald, G.~Vampa, P.~B.~Corkum, and T.~Brabec,
Phys.\ Rev.\ A \textbf{92}, 033845 (2015).

\bibitem{wu2}
M.~Wu, D.~A.~Browne, K.~J.~Schafer, and M.~B.~Gaarde,
Phys.\ Rev.\ A \textbf{94}, 063403 (2016).

\bibitem{ikemachi}
T.~Ikemachi, Y.~Shinohara, T.~Sato, J.~Yumoto, M.~Kuwata-Gonokami, and K.~L.~Ishikawa,
Phys.\ Rev.\ A \textbf{95}, 043416 (2017).
%
\bibitem{hansen}
K.~K.~Hansen, T.~Deffge, and D.~Bauer,
Phys.\ Rev.\ A \textbf{96}, 053418 (2017).
%

\bibitem{ikemachi2}
T.~Ikemachi, Y.~Shinohara, T.~Sato, J.~Yumoto, M.~Kuwata-Gonokami, and K.~L.~Ishikawa,
Phys.\ Rev.\ A \textbf{98}, 023415 (2018).
%

\bibitem{ikeda}
T.~N.~Ikeda, K.~Chinzei, and H.~Tsunetsugu,
Phys.\ Rev.\ A \textbf{98}, 063426 (2018). 

\bibitem{kruchinin2}
S.~Y.~Kruchinin,
arXiv:1806.05556.


\bibitem{kishida}
H.~Kishida, H.~Matsuzaki, H.~Okamoto, T.~Manabe, M.~Yamashita, Y.~Taguchi, and Y.~Tokura,
Nature \textbf{405}, 929 (2000).

\bibitem{ogasawara}
T.~Ogasawara, M.~Ashida, N.~Motoyama, H.~Eisaki, S.~Uchida, Y.~Tokura, H.~Ghosh, A.~Shukla, S.~Mazumdar, and M.~Kuwata-Gonokami,
Phys.\ Rev.\ Lett.\ \textbf{85}, 2204 (2000).





\bibitem{silva}
R.~E.~F.~Silva, I.~V.~Blinov, A.~N.~Rubtsov, O.~Smirnova, and M.~Ivanov, 
Nat.\ Photonics \textbf{12}, 266 (2018).  

\bibitem{murakami}
Y.~Murakami, M.~Eckstein, and P.~Werner, 
Phys.\ Rev.\ Lett.\ \textbf{121}, 57405 (2018). 

\bibitem{dejean}
N.~Tancogne-Dejean, M.~A.~Sentef, and A.~Rubio,
Phys.\ Rev.\ Lett.\ \textbf{121}, 97402 (2018).


\bibitem{takayoshi}
S.~Takayoshi, Y.~Murakami, and P.~Werner, 
Phys.\ Rev.\ B \textbf{99}, 184303 (2019).

\bibitem{zhu}
W.~Zhu, A.~Chacon, and J.~Zhu,
arXiv:1811.12334 

\bibitem{kino}
H.~Kino, and H.~Fukuyama, 
J.\ Phys.\ Soc.\ Jpn.\ \textbf{65}, 2158 (1996).

\bibitem{naka}
M.~Naka and S.~Ishihara,
J.\ Phys.\ Soc.\ Jpn.\ \textbf{79}, 063707 (2010).

\bibitem{hotta}
C.~Hotta, 
Phys.\ Rev.\ B \textbf{82}, 241104(R) (2010).

\bibitem{tsuchiizu}
M.~Tsuchiizu (private communication).


\bibitem{MM}
Derivation of the effective models and introduction of the light field in the models are explained in more detail in the Supplemental Material.

\bibitem{kirilyuk}
P.~Pfeuty, 
Ann.\ Phys.\ \textbf{57}, 79 (1970). 

\bibitem{vidal}
G.~Vidal, Phys.\ Rev.\ Lett.\ \textbf{98}, 070201 (2007).


\bibitem{raedt}
H.~De~Raedt, S.~Miyashita, K.~Saito, D.~Garc\'ia-Pablos and N.~Garc\'ia, 
Phys.\ Rev.\ B \textbf{56}, 11761 (1997). 

\bibitem{miyashita}
S.~Miyashita, K.~Saito, and H.~De~Raedt, 
Phys.\ Rev.\ Lett.\ \textbf{80}, 1525 (1998). 


\bibitem{javadi}
H.~H.~S.~Javadi, R.~Laversanne, and A.~J.~Epstein, 
Phys.\ Rev.\ B \textbf{37}, 4280 (1988). 

\bibitem{monceau}
P.~Monceau, F.~Ya.~Nad, and S.~Brazovskii, 
Phys.\ Rev.\ Lett.\ \textbf{86}, 4080 (2001).

\bibitem{iwai}
Y.~Naitoh, Y.~Kawakami, T.~Ishikawa, Y.~Sagae, H.~Itoh, K.~Yamamoto, T.~Sasaki, M.~Dressel, S.~Ishihara, Y.~Tanaka, K.~Yonemitsu, and S.~Iwai, 
Phys.\ Rev.\ B \textbf{93}, 165126 (2016).

\end{thebibliography}
\end{document}


\title{Supplemental Material for ``High-Harmonic Generation in a Correlated Electron System''}  
\author{Shohei Imai}
\author{Atsushi Ono}
\author{Sumio Ishihara}
\affiliation{Department of Physics, Tohoku University, Sendai 980-8578, Japan}

\maketitle
\narrowtext

\section{Effective Hamiltonian}
In this section, we discuss the effective Hamiltonians of the interacting electron system in a dimer-type lattice structure. 
We start from the extended Hubbard model on a one-dimensional dimer-type lattice given by 
\begin{align}
{\cal H}&=
-\sum_{i s} \left(t_0 c_{i a s}^{\dagger}c_{i bs}+{\rm H.c.} \right)
\nonumber \\
&\quad -\sum_{i s} \left(t' c_{i b s }^{\dagger}c_{i+1 a s }+{\rm H.c.} \right)
\nonumber \\
&\quad +U\sum_{i \gamma} {\tilde n}_{i \gamma \uparrow} {\tilde n}_{i \gamma \downarrow} \nonumber \\
&\quad +V_0\sum_i {\tilde n}_{i a} {\tilde n}_{i b} + V'\sum_i {\tilde n}_{i b} {\tilde n}_{i+1 a} ,
\label{eq:extHub}
\end{align}
where $c_{i \gamma s}^\dagger$ ($c_{i \gamma s}$) 
is the creation (annihilation) operator of an electron at site the $i$th unit cell and sublattice $\gamma\ (=a, b)$ with spin $s\ (= {\uparrow}, {\downarrow})$, 
and ${\tilde n}_{i \gamma}=\sum_{s}{\tilde n}_{i \gamma s}=\sum_s c_{i \gamma s}^\dagger c_{i \gamma s}$ is the number operator. 
The first two terms represent the electron hoppings between the nearest neighbor (NN) sites, the third term describes the on-site Coulomb interaction, and the last two terms describe the long-rang Coulomb interactions between the NN sites. 
The electron number per site is set to 0.5. 

In the limit of large $U$ in comparison with other energy parameters, it is expected that the electron occupancy in each site is less than one, and the spin degree of freedom does not play essential role in the optical processes.  
Then, the electron operators are replaced by the spinless-fermion operators, the third term in Eq.~(\ref{eq:extHub}) is neglected, and thus the effective Hamiltonian is given by 
\begin{align}
{\cal H}_{\rm SF}&=
-\sum_i \left( t_0 f_{i a}^{\dagger} f_{i b} + {\rm H.c.} \right) 
-\sum_i \left( t' f_{i b}^{\dagger} f_{i+1 a} + {\rm H.c.} \right) \nonumber \\
&\quad +V_0\sum_i n_{i a} n_{i b} + V'\sum_i n_{i b}n_{i+1 a} ,
\label{eq:sfM}
\end{align}
which is Eq.~(1) in the main text (MT). 
We introduce the the creation (annihilation) operator of a spinless fermion  $f_{i \gamma}^{\dagger}$ $(f_{i \gamma})$ at the $i$th unit cell and sublattice $\gamma$, and 
the number operator for a spinless fermion $n_{i \gamma}=f_{i \gamma}^\dagger f_{i \gamma}$. 
The number of the spinless fermion per site is 0.5.

In the limit of the strong dimerization, i.e., $t' \ll t_0, V_0$, 
it is convenient to describe the electronic states inside a dimer unit by using the pseudo-spin (PS) operator with magnitude of $1/2$.  We introduce the PS operator as 
\begin{align}
{\bm \sigma}_i=\sum_{\gamma \gamma'}  f_{i \gamma}^\dagger {\bar {\bm \sigma}}_{\gamma \gamma'} f_{i \gamma'} ,  
\end{align}
with the Pauli matrices $\bar {\bm \sigma}$. 
%
The intra-dimer Coulomb interaction term is replaced by a constant, and the inter-dimer one is transformed into 
$-V'/4 \sum_{i} \sigma_{i}^z \sigma_{i+1}^z$. 
Although additional inter-dimer interactions are induced by the second order processes with respect to $t'$, 
we neglect these terms for simplicity. 
Then, we have the transverse Ising (TI) model as an effective model of Eq.~(\ref{eq:sfM}) 
given as~\cite{tsuchiizu}, 
\begin{align}
{\cal H}_{\rm TI}=-\frac{V'}{4} \sum_i \sigma_i^z \sigma_{i+1}^z -t_0 \sum_i \sigma_i^x ,  
\label{eq:psM}
\end{align}
which is Eq.~(2) in MT.

In the light field, the vector potential $A(t) $ is introduced in Eq.~(\ref{eq:sfM}) as the Peirerls phase given as 
$t_0 \to t_0 e^{-iA(t)}$ and $t' \to t' e^{-iA(t)}$, where a difference of the bond lengths is neglected. 
In the TI model, in which the linear term of $t'$ is neglected, 
the coupling with the vector potential is introduced by replacing the second term in Eq.~(\ref{eq:psM}) as 
\begin{align}
 -t_0 \sum_i {\left[ \cos A(t) \sigma_i^x-\sin A(t) \sigma_i^y \right ]} .
\label{eq:eA}
\end{align}
 When we consider the tunneling phenomena, it is useful to consider the response to the electric field $E(t)$ instead of the vector potential. 
 The coupling with the electric field is introduced in the TI model as 
 \begin{align}
- \frac{E(t)}{2} \sum_i \sigma_i^z , 
\label{eq:eE}
 \end{align}
 where $(1/2) \sum_i \sigma_i^z$ is identified as the electric dipole moment inside of dimer units.

The electric current operator is identified as 
\begin{align}
\hat{j}(t) &= -\frac{1}{N}\frac{\delta{\cal H}}{\delta A(t)} \nonumber \\
&= -t_0 \frac{1}{N}\sum_{i} {\left [ \cos A(t) \sigma_i^x + \sin A(t) \sigma_i^y \right ]},
\end{align}
where Eq.~(\ref{eq:eA}) is adopted as the coupling term with light. 
This vector-potential picture represented by the wavefunction $|\psi (t) \rangle$ is changed into the electric-field picture represented by $|\psi' (t) \rangle $ where Eq.~(\ref{eq:eE}) is adopted. 
By introducing the unitary transformation  
 $|\psi'(t)\rangle=U(t)|\psi(t) \rangle$ with $U(t)=\exp[-iA(t) \sum_i \sigma_i^z/2]$, 
the current operator in the electric-field picture is given by 
\begin{align}
\hat{j}' = -t_0\frac{1}{N}\sum_i \sigma_i^y ,
\end{align}
which satisfies the relation $\langle\psi'(t)|\hat{j}'|\psi'(t)\rangle=\langle \psi(t)|\hat{j}|\psi(t)\rangle$.

\begin{figure}[t]
\centering
\includegraphics[width=\columnwidth, clip]{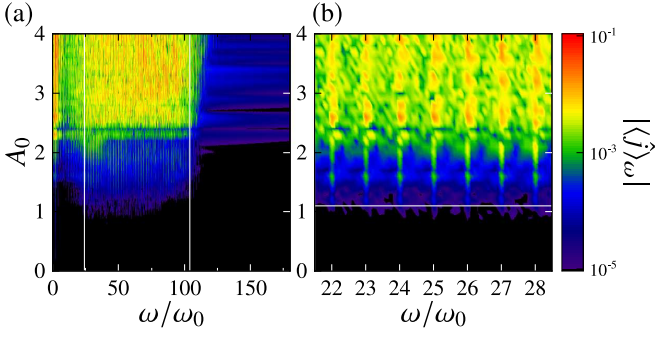}
\caption{
(a) Intensity maps of $\langle \hat{j} \rangle_\omega$ in the $\omega$-$A_0$ plane in the case of $V'/(4t_0)=1.6$. 
The vertical lines show the lower and upper edges of the kink-antikink bands. 
(b) An enlargement of (a) around $21<\omega/\omega_0<29$. 
The horizontal line indicates threshold intensity. 
Parameter values are chosen to be $\omega_0/t_0=0.1$, and $\tau=1/\omega_0$. 
The iTEBD method with $\chi=200$ is utilized. 
}
\label{fig:fig1}
\end{figure}
\begin{figure}[t]
\centering
\includegraphics[width=\columnwidth, clip]{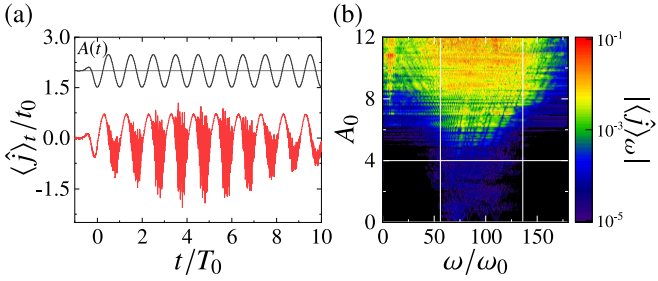}
\caption{
Calculated results obtained by the ED method in the cw field. 
(a) Time profiles of $A(t)$ (upper) and $\langle j \rangle_t$  (lower), and (b) an intensity map of the HHG spectra in the $\omega$-$A_0$ plane. 
The vertical lines and a horizontal line in (b) represent the lower- and upper-edges of the kink-antikink band, and threshold intensity, respectively. 
Parameter values are chosen to be $V'/(4t_0)=2.4$, $\omega_0/t_0=0.1$, $\tau=1/\omega_0$, and $N=18$. 
}
\label{fig:fig2}
\end{figure}

\section{Interaction parameter dependence}
In Fig.~\ref{fig:fig1}, the Fourier transforms of the currents $\langle \hat{j} \rangle_\omega$ in the case of $V'/(4t_0)=1.6$ are presented. 
The infinite time-evolving block decimation (iTEBD) method with $\chi=200$ is utilized to analyze the TI mode in the continuous-wave (cw) field. 
The threshold behavior of the HHG spectra with respect to amplitude of the vector potential, $A_0$,  are commonly observed, and 
the threshold increases with increasing $V'$.  

\section{Finite size effects}
In order to confirm the calculated results by the iTEBD method shown in MT, we show the results obtained by the exact diagonalization (ED) method based on the Lanczos algorithm. 
A periodic boundary condition is imposed on an $N$-site cluster. 
We introduce the cw light.  
A time profile of the induced current $\langle \hat{j} \rangle_t$ and an intensity plot of the Fourier transform of current $\langle \hat{j} \rangle_\omega$ calculated by the ED method are shown in Figs.~\ref{fig:fig2}(a) and \ref{fig:fig2}(b), respectively. 
Overall features of $\langle \hat{j} \rangle_t$ and $\langle \hat{j} \rangle_\omega$ almost reproduce the results shown in Figs.~1(d) and 2(a) in MT. 
We confirm the following two characteristics in $\langle \hat{j} \rangle_\omega$: 
the HHG spectra show a plateau mainly between the upper and lower edges of the kink-antikink excitation spectra, i.e., 
$50 \lesssim \omega/\omega_0 \lesssim 130$, and there is a threshold $A_0 \sim 4$ with respect to amplitude of light.


\begin{figure}[t]
\centering
\includegraphics[width=\columnwidth, clip]{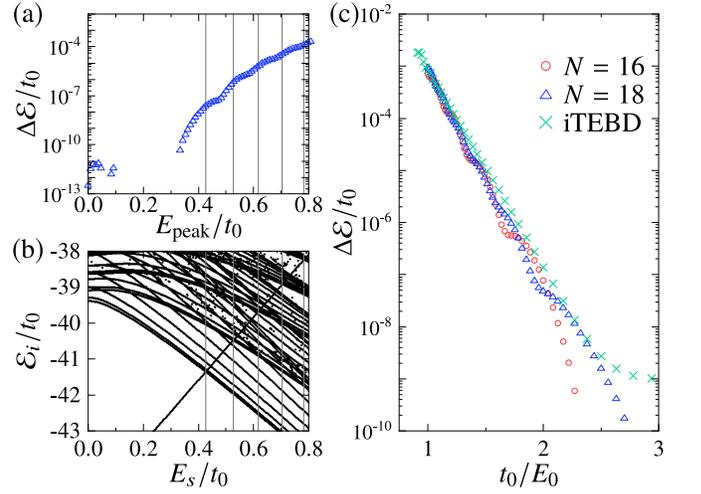}
\caption{
Calculated results obtained by the ED method in the one-cycle pulse field. 
(a) Absorbed energy $\Delta {\cal E}$ as a function of peak height of the pulse electric field $E_{\rm peak}$. 
Data below $E_{\rm peak}=0.1$ is attributed to calculation artifact. 
(b) Energy levels as functions of the static electric field $E_s$. 
The vertical lines represents the fields in which the state, which is adiabatically connected to the ground state anticrosses with the excited states. 
(c) Absorbed energy $\Delta {\cal E}$ as a function $1/E_0$. 
Results in $N=16$ and $18$ are plotted by circles and triangles, respectively. 
Results calculated by the iTEBD method with the truncation parameter $\chi=200$ are shown by crosses. 
Parameter values are chosen to be $V'/(4t_0)=2.4$, $\omega_0/t_0=0.1$, and $\tau=1/\omega_0$. 
Cluster size in (a) and (b) is $N=18$. 
}
\label{fig:fig3}
\end{figure}

Results for the one-cycle pulse calculated by the ED method are shown in Fig.~\ref{fig:fig3}, corresponding to 
Fig.~2{(c)} in MT calculated by the iTEBD method. 
The absorbed energy after the pulse irradiation $\Delta {\cal E}$
is plotted as a function of the electric field $E_{\rm peak}$ in Fig.~\ref{fig:fig3}(a). We define 
peak height of the electric field  
$E_{\rm peak}=0.928E_0$ with $E_0=A_0 \omega_0$. 
The exponential-like behavior mentioned in MT, i.e., $\Delta {\cal E} \propto \exp(-E_{\rm th}/E_{0})$ with the threshold electric field $E_{\rm th}$, is shown. 
Calculated data around $E_{\rm peak} \lesssim 0.1$ are attributed to the numerical artifact. 
Weak multiple dip-and-hump structures appear in the $\Delta {\cal E}$-$E_{\rm peak}$ curve.
The energy level diagram under the static electric field $E_s$ is shown in  Fig.~\ref{fig:fig3}(b), which is a part of 
Fig.~3(a) in MT. 
Each dip-and-hump structure corresponds to an anticrossing point between ${\cal E}_0$, which is adiabatically connected to the ground state at $E_s=0$ and an excited energy level, as indicated by thin vertical lines. 
This implies that these structures originate from Landau-Zener tunneling processes.
The absorbed energies calculated by the ED method are plotted as function of $1/E_0$ in Fig.~\ref{fig:fig3}(c) for $N=16$ and $18$.  
The dip-and-hump structure is weakened with increasing $N$, and the data by the ED method tends to converge to that by the iTEBD method. 
Calculated data sets almost reproduce the results obtained by the iTEBD method where the truncation parameter is chosen to be $\chi=200$. 

\section{Optical absorption spectra}
\begin{figure}[t]
\centering
\includegraphics[width=0.7\columnwidth, clip]{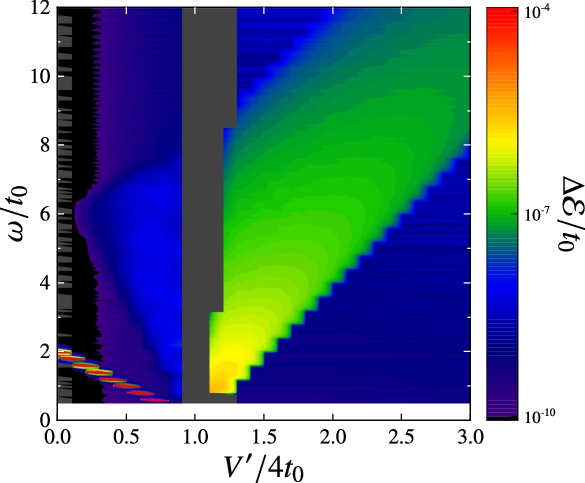}
\caption{
Optical absorption spectra calculated by the iTEBD method. 
Color represents amplitude of the energy increment after the irradiation. 
Gray area represents a parameter region where the data with enough accuracy are not obtained in the iTEBD method.  
We set $\chi=40$. 
}
\label{fig:fig4}
\vspace{5mm}
\centering
\includegraphics[width=\columnwidth, clip]{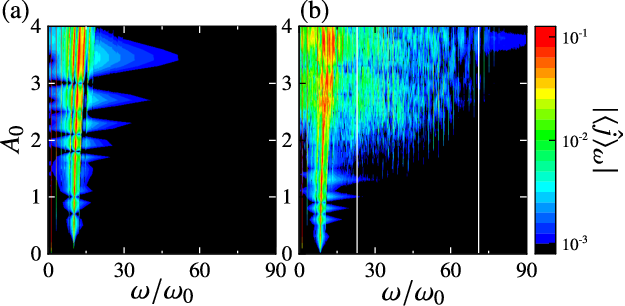}
\caption{
Intensity maps of the Fourier transforms of current calculated in the DM phase. 
Results in (a) and (b) are obtained in the TI model and the spinless-fermion model, respectively. 
Parameter values are chosen to be $V'/(4t_0)=0.5$, $\omega_0/t_0=0.1$,  $\tau=1/\omega_0$, $t'/t_0=0.5$, and $V_0/V'=1$. 
Horizontal lines in (b) represent the edges of the optical absorption spectra. 
}
\label{fig:fig5}
\end{figure}

The optical absorption spectra are calculated by the iTEBD method. 
The energy increment after the weak irradiation ($A_0=0.0001$) is calculated in the real time evolution. 
The intensity map of the spectra is shown in Fig.~\ref{fig:fig4} which reproduces the previously known results schematically shown in Fig.~1(b) in MT, except for the region near $V'/(4t_0)=1$ where the data with enough accuracy are not obtained in the iTEBD method.

\vspace{2\baselineskip}
\section{HHG in dimer-Mott phase}

In order to compare the HHG spectra in the polar CO state shown in MT, we show here the results of HHG in the dimer-Mott (DM) state. 
Intensity maps of the Fourier transforms of the current $\langle \hat{j} \rangle_\omega$ are calculated in the DM state. 
Results in the TI model and the spinless-fermion model are shown in Figs.~{\ref{fig:fig5}}(a) and \ref{fig:fig5}(b), respectively.  
The current in $\omega/\omega_0 \approx 10$ and around $20 \lesssim \omega/\omega_0 \lesssim 70$  
originates from the exciton excitation and the individual excitation of fermions, respectively.  
Owing to the space inversion symmetry, 
the HHG appears at $\omega=n \omega_0$ where $n$ is odd numbers . 
In the TI model, the HHG signal due to the exciton only appears, since the electrons are localized inside of the dimer units.


